\documentclass[a4paper,final]{appolb}
\usepackage{graphicx}
\usepackage{epsfig}
\usepackage{mathrsfs}
\usepackage[compress]{cite} 
\usepackage[hypertex]{hyperref}
\hypersetup{colorlinks,citecolor=blue,linkcolor=blue}

\begin{document}
\title{Importance of volume corrections on the net-charge distributions at the RHIC BES energies%
	\thanks{Presented at CPOD 2016: Critical Point and Onset of Deconfinement 2016, Wroclaw, Poland, May 30th - June 4th, 2016}%
}
\author{Hao-jie Xu\thanks{haojiexu@pku.edu.cn}
	\address{Department of Physics and State Key Laboratory of Nuclear Physics
	and Technology, Peking University, Beijing 100871, China}
}
\maketitle
\begin{abstract}
	The paper presents my recent investigations of volume corrections on the cumulant 
	products of net-charge distributions in statistical model, corresponding to the data 
	reported by the STAR collaboration. The corrected statistical expectations, under
	simple Poisson approximations, can reasonably explain the data 
	measured in experiment. The results indicate that volume corrections play crucial 
	role in event-by-event multiplicity fluctuation studies. 
\end{abstract}

\section{Introduction}
The multiplicity fluctuations 
with event-by-event analysis are expected to provided us some crucial
information about the critical end point of Quantum Chromodynamics phase diagram in the $(T,\mu_{B})$ plane~\cite{Jeon:1999gr,Stephanov:2008qz,Bazavov:2012vg,Gupta:2011wh,Chen:2015dra,Jiang:2015hri}.
Some observables, e.g. the cumulants of (net-conserved) charge distributions,
have been measured by the beam energy scan (BES) program  
from the Relativistic Heavy Ion Collider (RHIC) at Brookhaven National 
Laboratory (BNL) with a wide range of collision energies  from 
$\sqrt{s_{NN}}=7.7$ GeV to $\sqrt{s_{NN}}=200$ GeV \cite{Aggarwal:2010wy,Adamczyk:2013dal, Adamczyk:2014fia}. 

Besides the fluctuation data and theoretical studies on critical fluctuations, 
it is clear that a sufficient understanding of non-critical statistical fluctuations is also
important.
To bridge the gaps between experimental measurements and statistical fluctuation calculations,
I have derived a general formalism in statistical model~\cite{Xu:2016jaz}
for recent  multiplicity fluctuation measurements at RHIC~\cite{Aggarwal:2010wy,Adamczyk:2013dal, Adamczyk:2014fia}.
With the volume corrections, the multiplicity distributions can be written as~\cite{Xu:2016jaz,Xu:2016qzd}
\begin{equation} 
	\mathscr{P}_{B|A}(q|k) = \frac{1}{\mathscr{P}_{A}(k)}\int dV F(V)P_{A}(k;V)P_{B}(q;V),
	\label{eq:distribution}
\end{equation}
where $\mathscr{P}_{A}(k)$ is the distribution of reference multiplicity
\begin{equation} 
	\mathscr{P}_{A}(k) = \int dV F(V)P_{A}(k;V).\label{eq:genrefmultiplicity} 
\end{equation}
Here $q$ represents the multiplicity of fluctuation measures
in moment analysis sub-event $B$, and $k$ represents the multiplicity
of reference particles  in centrality definition sub-event $A$. 
$P_{A}(k;V)$ and $P_{B}(q;V)$ stand for multiplicity distributions in a fixed volume $V$. 
Here I have assumed that the two sub-events are 
independent of each other in each event.

\section{Corrected cumulant products of net-charge distributions}
First I assume that the probability distribution $P_{A}(k;V)$ of reference particles in a fixed volume
is Poisson distribution. Then the volume V can be substituted by $\lambda$, the Poisson parameters of $P_{A}(k)$.
At non-central collisions, the first four cumulants of $\mathscr{P}_{B}(q|k)$ read~\cite{Xu:2016qzd, Skokov:2012ds} 
\begin{eqnarray}
	\label{eq9}
	\frac{c_{1}}{\langle\lambda\rangle}&=&\kappa_{1}\nonumber\\
	\frac{c_2}{\langle\lambda\rangle} &=& \kappa _2 + \kappa _1^2 , \nonumber\\
	\frac{c_3}{\langle\lambda\rangle} &=& \kappa _3+3 \kappa _2 \kappa _1 + 2 \kappa _1^3 , \nonumber \\
	\frac{c_4 }{\langle\lambda\rangle}&=& \kappa _4 +4 \kappa _3 \kappa _1+3 \kappa _2^2 +
	12 \kappa _2 \kappa _1^2  + 6 \kappa _1^4, 
\end{eqnarray}
where $\kappa_{1} ,\kappa_{2},\kappa_{3}$ and $\kappa_{4}$ are the first four reduced cumulants of $P_{B}(q;\lambda)$~\cite{Xu:2016qzd} 
\begin{eqnarray}
	\kappa_{1} &=& \frac{\sum q P_{B}(q;\lambda)}{\lambda}\equiv \frac{\bar{q}}{\lambda}, \nonumber \\
	\kappa_{2} &=& \frac{\sum (\Delta q)^{2} P_{B}(q;\lambda)}{\lambda}, \nonumber \\
	\kappa_{3} &=& \frac{\sum (\Delta q)^{3} P_{B}(q;\lambda)}{\lambda},   \\
	\kappa_{4} &=& \frac{ \sum(\Delta q)^{4} P_{B}(q;\lambda) - 3 (\sum (\Delta q)^{2}P_{B}(q;\lambda))^{2}}{\lambda}, \nonumber
\end{eqnarray}
with $\Delta q\equiv q - \bar{q}$ and 
\begin{equation}
	\langle\lambda\rangle = (k+1)\frac{\mathscr{P}_{A}(k+1)}{\mathscr{P}_{A}(k)}\simeq k+1.  
	\label{eq:lambdam}
\end{equation}

If the net-charge distributions are Skellam distributions in a fixed volume, one obtains $\kappa_1=\kappa_3 = \kappa _+ - \kappa _-$ 
and $\kappa_2=\kappa_4=\kappa _+ + \kappa_ -$, where $\kappa_\pm = M_\pm / (k+1) $  and 
$M_\pm$ are the mean value of positive and negative charges. The cumulant products of net-charge distributions can be written as ~\cite{Xu:2016jaz,Xu:2016qzd}
\begin{eqnarray}
	\omega & \equiv & c_2/c_1 = \beta(1-\alpha)+\frac{1+\alpha}{1-\alpha}, \nonumber  \\
	S\sigma & \equiv&  c_3/c_2= 2\beta(1-\alpha)+\frac{\beta(1-\alpha^{2})+1-\alpha}{\beta(1-\alpha)^{2}+1+\alpha}, \nonumber \\
	\kappa\sigma^{2}& \equiv & c_4/c_2 = 6\beta(\gamma-\frac{2\alpha}{\gamma})+1, \label{eq:cumulantp}
\end{eqnarray}
with $\alpha= \kappa_{-}/\kappa_{+}=M_-/M_+$, $\beta = \kappa_{+} = M_{+}/(k+1)$, $\gamma=\beta(1-\alpha)^{2}+1+\alpha$.
In experiment~\cite{Adamczyk:2014fia}, the kinematic cut for the reference
particles (total charges) in sub-event $A$ is $1.0>|\eta|>0.5$ and for the fluctuation measures (net-charges)
in sub-event $B$ is $|\eta|<0.5$, where $\eta$ is pseudorapidity. Therefore,
due to the multiplicity distributions as function of $\eta$ are almost platform-like distributions at mid-rapidity, one obtains
$M_{+}+M_{-}\simeq k+1$ and $\beta\simeq 1/(1+\alpha)$. Eq.~(\ref{eq:cumulantp}) can be written as
\begin{eqnarray}
	\omega & \simeq & \frac{1-\alpha}{1+\alpha}+\frac{1+\alpha}{1-\alpha}\simeq \frac{1+\alpha}{1-\alpha}, \nonumber  \\
	S\sigma & \simeq&  2\frac{1-\alpha}{1+\alpha} + \frac{2(1-\alpha)}{\frac{(1-\alpha)^{2}}{1+\alpha}+1+\alpha} \simeq 4\frac{1-\alpha}{1+\alpha}, \nonumber \\
	\kappa\sigma^{2}& \simeq & \frac{6(\gamma-\frac{2\alpha}{\gamma})}{1+\alpha}+1\simeq 7 - \frac{12\alpha}{(1+\alpha)^{2}} \simeq 4, \label{eq:cumulantp2}
\end{eqnarray}
where I have used $\alpha\simeq 1$. 

\begin{figure}
	\begin{center} 
		\includegraphics[scale=0.33]{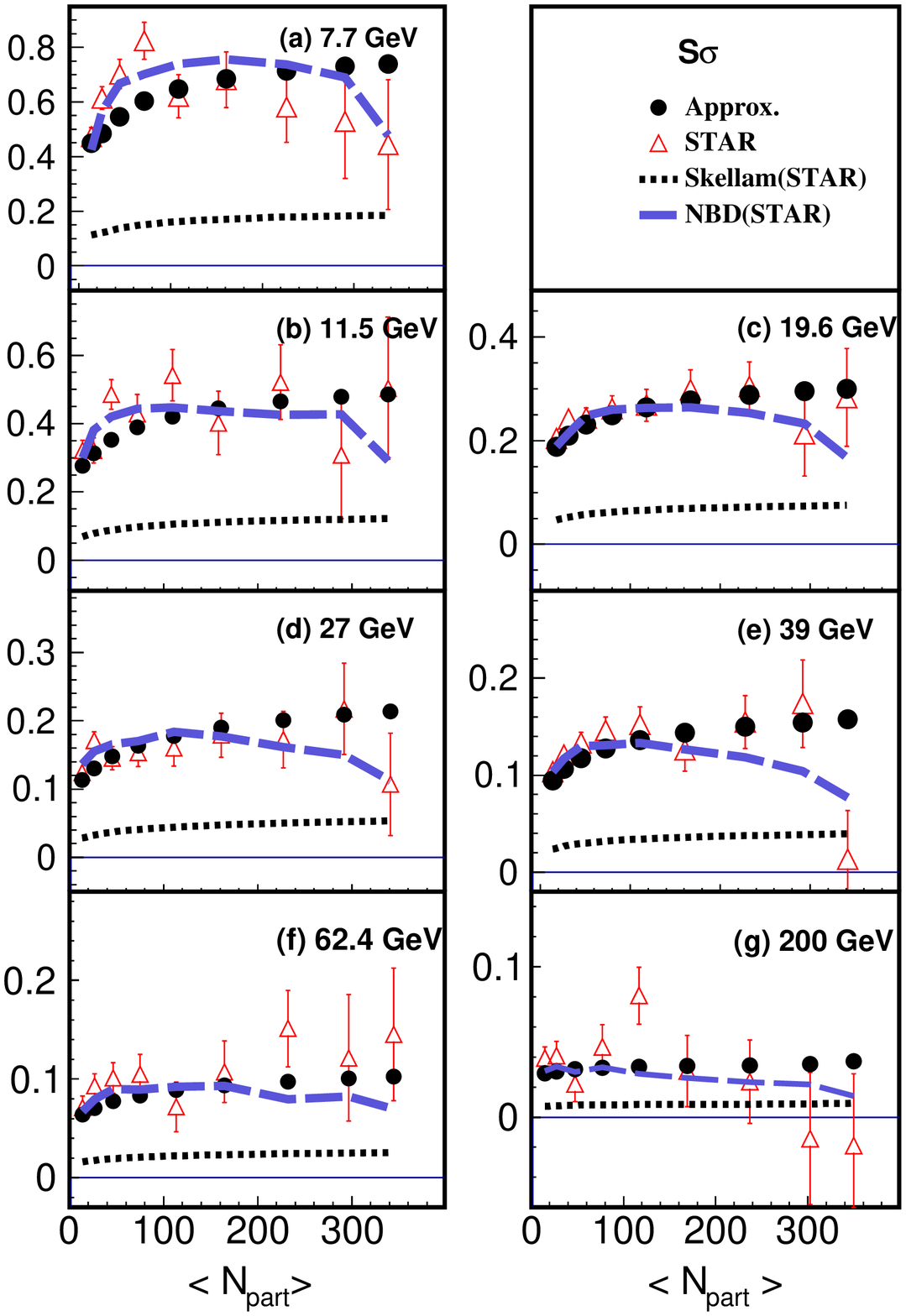} \includegraphics[scale=0.33]{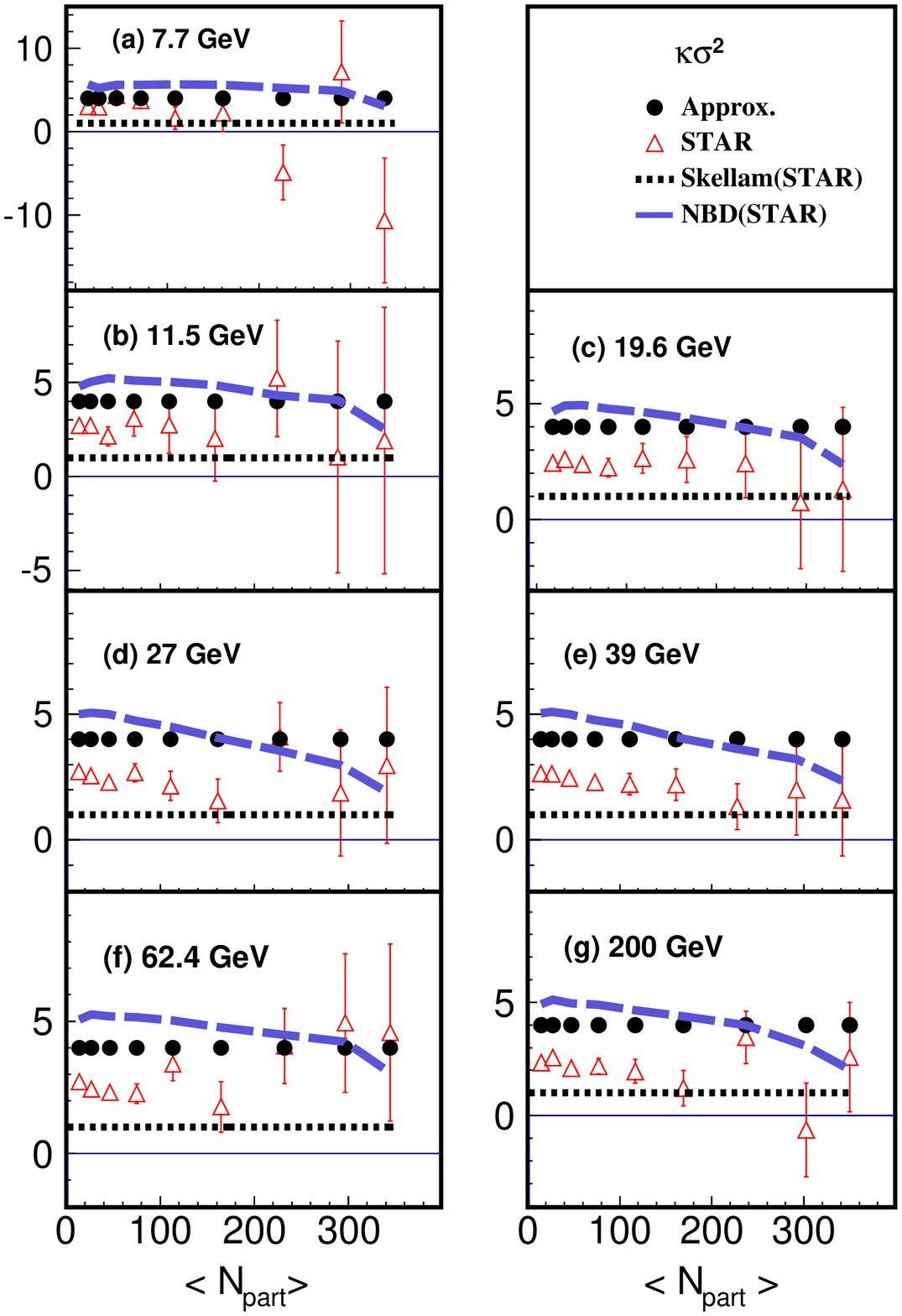}
	\end{center}
	\caption{(Color online). Volume-corrections-corrected $S\sigma$ (left) and $\kappa\sigma^2$(right) of the
		net-charge multiplicity distribution in Au+Au collisions at
		$\sqrt{s_{NN}}=7.7$ to $200$ GeV.  The data, Skellam and NBD
		baselines are taken from ~\cite{Adamczyk:2014fia}. 
	\label{fig:SsigmaBES}} 
\end{figure}

In Eq.~(\ref{eq:cumulantp2}), the scale variance $\omega$ is close to the Skellam expectation,
which indicate that the volume corrections on scale variances of net-charge distributions 
can be neglected. However, the values of $S\sigma$ and $\kappa\sigma^{2}$ of net-charge
distributions are about four times of the Skellam expectations. The corrected cumulant products
and data are shown in Fig.~\ref{fig:SsigmaBES}. The corrected $S\sigma$ are 
closer to the experiment data/NBD baselines than the Skellam baselines given 
in~\cite{Adamczyk:2014fia}. The corrected $\kappa\sigma^{2}$ are close to
the NBD baselines, but fail to quantitatively reproduce the data. This indicate 
the existence of correlations between positive and negative charges~\cite{Adamczyk:2014fia} 
and/or the correlations between the fluctuation measures and the reference particles.
The results indicate that the volume corrections play crucial role for the data 
of net-charge distributions reported by the STAR collaboration~\cite{Adamczyk:2014fia}.

Note that, though it have been shown in the figures, the results 
in $0-5\%$ centrality bins are questionable. This is due to 
the non-trivial features of volume distributions at most-central collisions, 
which are reflected in the rapid decreasing of reference multiplicity 
distributions at top few centrality percentage. With Glauber volume 
distributions, I found that
the non-critical volume corrections on high order cumulants become weak 
at most-central collisions~\cite{Xu:2016qzd}. However, the details of volume distribution
in relativistic heavy ion collisions are required for more 
precise event-by-event multiplicity studies.

\section{Conclusions and Discussions}
The volume corrections on net-charge distributions at non-central heavy 
ion collisions have been investigated. The multiplicity fluctuations of reference particles and fluctuation measures
in a fixed volume are simulated by Poisson and Skellam  distributions.  
The volume corrections make significant contribution to the 
measured cumulants products $S\sigma$ and $\kappa\sigma^{2}$ of net-charge distributions 
reported by the STAR collaboration, but can be neglected for the scale variance $\omega$. 

Note that, even in a fixed volume, there are
many other effects that make the multiplicity distributions deviate
from Poisson distributions. These corrections,
e.g., finite volume effect, quantum effect,
resonance decays, experimental acceptance, etc~\cite{Becattini:1995xt,Cleymans:2004iu,BraunMunzinger:2011ta},
should be taken into account, especially in the case of net-proton distributions reported by the STAR collaboration. 
\newline

This work is supported by the China Postdoctoral Science Foundation under
grant No.~2015M580908.

\end{document}